\documentclass[a4paper]{jpconf}
\usepackage{graphicx}
\begin{document}
\title{Functional approach for pairing in finite systems: 
{\it How to define restoration of broken symmetries in Energy Density Functional theory ?}}





\author{G. Hupin and D. Lacroix}
\address{
Grand Acc\'el\'erateur National d'Ions Lourds (GANIL), CEA/DSM-CNRS/IN2P3, Bvd Henri Becquerel, 14076 Caen, France}

\author{M. Bender}
\address{Universit{\'e} Bordeaux,
             Centre d'Etudes Nucl{\'e}aires de Bordeaux Gradignan, UMR5797,
             F-33175 Gradignan, France,  \\
             CNRS/IN2P3,
             Centre d'Etudes Nucl{\'e}aires de Bordeaux Gradignan, UMR5797,
             F-33175 Gradignan, France
             }
             
\ead{hupin@ganil.fr, lacroix@ganil.fr, bender@ganil.fr}

\begin{abstract}
The Multi-Reference Energy Density Functional (MR-EDF) approach (also called configuration mixing or Generator Coordinate Method), that is commonly used to treat pairing in finite 
nuclei and project onto particle number, is re-analyzed. 
It is shown that, under certain conditions, the MR-EDF energy can be interpreted as a functional of the one-body 
density matrix of the projected state with good particle number. Based on this observation, we propose a new approach, called 
Symmetry-Conserving EDF (SC-EDF), where the breaking and restoration of symmetry are accounted for simultaneously. We show, that such an approach is free from pathologies recently observed in MR-EDF and can be used with a large flexibility on the density dependence of the functional.
\end{abstract}

\section{Introduction}
In the last decade, configuration mixing has become a standard tool in the Energy Density Functional approach 
both to restore symmetries and to provide a competitive theory for nuclear spectroscopy. With the possibility to perform 
systematic and precise studies, detailed analysis of MR-EDF have been made, pointing out some questionable aspects related 
to the 
way MR-EDF is constructed \cite{Ang01,Dob07,Lac09,Ben09,Dug09}. Below, we further investigate 
the MR-EDF applied to particle number restoration. The Hamiltonian case is first discussed 
as well as some specific features that will be useful for the EDF case. 
\section{Particle number restoration in the Hamiltonian case}
Starting from a two-body Hamiltonian, written here as:
\begin{equation}
\label{eq:hamil} H =   \sum_{ij} t_{ij} \, a^\dagger_i \, a_j + \frac{1}{4} \sum_{ijkl} \bar{v}_{ijkl} \, a^\dagger_i \, a^\dagger_j \, a_l \, a_k
\, \, \, ,
\end{equation}
where $\bar{v}$ denote the anti-symmetric two-body interaction matrix elements,
pairing can be treated approximately by minimizing the energy in the space of quasi-particle vacua. For a given quasi-particle 
reference state of the special Bogolyubov type, denoted by $| \Phi_0 \rangle$, the energy writes  
 \begin{eqnarray}
 { E}_{SR}[\Phi_0] =&&{E}_{SR} \left[ \rho , \kappa, \kappa^* \right] =  \sum_{i} t_{ii} \rho_{ii}+ 
 \frac{1}{2} \sum_{i,j} \overline{v}_{ijij}^{}  \rho_{ii}\rho_{jj}  
+ \frac{1}{4}  \sum_{i,j} \overline{v}_{i\bar\imath j\bar\jmath }^{} \kappa_{i \bar\imath }^* \kappa_{j \bar\jmath } \, ,
\label{eq:denssrhamil}
\end{eqnarray} 
where $\rho$ and $\kappa$ denote the normal and anomalous density respectively and where the set of single-particle states are those where $\rho$ is diagonal, i.e. the so-called canonical basis of $\rho$.  For non-zero anomalous density, the 
U(1) symmetry is explicitly broken and the particle number is only imposed in average.

In a second step, this symmetry can be restored by projecting out the component with $N$
particles $| \Psi_N \rangle = P^N | \Phi_0 \rangle $ where $P^N$ denotes the particle number projection operator defined through\cite{Rin80}
\begin{equation}
\label{eq:Pop}
{P}^N
=  \frac{1}{2\pi} \int_{0}^{2\pi} \! d{\varphi} \; \,e^{i\varphi (\hat{N}-N)}
\, .
\end{equation}

The expectation value of any operator $O$ that conserves particle number can then be 
expressed as 
\begin{equation}
\label{eq:expproj}
\frac{\langle \Psi_{N} | \, O \, | \Psi_{N} \rangle}{\langle \Psi_{N} |  \Psi_{N} \rangle}
=  \int_{0}^{2\pi}  d\varphi \, \frac{
\langle \Phi _0 | O  | \Phi_\varphi \rangle}{
\langle \Phi _0  | \Phi_\varphi \rangle}   {\cal N}_N ({0, \varphi}) 
\, ,
\end{equation} 
where the shorthand  
\begin{eqnarray}
{\cal N}_N ({0, \varphi}) \equiv  \frac{e^{-i\varphi N}}{2\pi}
 \frac{\langle  \Phi_0 | \Phi _{\varphi} \rangle}{\langle \Psi_{N} |  \Psi_{N} \rangle}\, ,
\end{eqnarray}
has been introduced. Here $\varphi$ denotes the gauge angle, whereas 
$| \Phi_{\varphi} \rangle = e^{i \varphi \hat N} \, | \Phi_0 \rangle$ refers to the state  $| \Phi_0 \rangle$ 
rotated in gauge space. For $O = H$, an expression is obtained for the projected energy: 
\begin{eqnarray}
{E}_N [\Psi_{N}] \equiv   \int_{0}^{2\pi}  d\varphi \, {E}_{SR} \left[ \rho^{0\varphi} , \kappa^{0\varphi}, 
{\kappa^{\varphi 0 }}^\star \right]
   {\cal N}_N ({0, \varphi}) \, .  \label{eq:ekernel}
\end{eqnarray}
The energy kernel are defined through
\begin{eqnarray}
{E}_{SR} \left[ \rho^{0\varphi} , \kappa^{0\varphi}, {\kappa^{\varphi 0 }}^\star \right] &\equiv&  \frac{\langle  \Phi_0 | H | \Phi _{\varphi} \rangle }{\langle  \Phi_0 | \Phi _{\varphi} \rangle
} . 
\end{eqnarray}

An equivalent form of the projected energy $E_N$ can directly be found by taking the expectation value of 
$H$ with the projected state, leading to:
\begin{eqnarray}
{E}_{N} \left[ \Psi_N \right] & = &  \frac{\langle  \Psi_N | H | \Psi _{N} \rangle }{\langle  \Psi_N | \Psi _{N} \rangle }=
   \sum_{ij} t_{ij} \,  \frac{\langle  \Psi_N |    a^\dagger_i \, a_j  | \Psi _{N} \rangle }{\langle  \Psi_N | \Psi _{N} \rangle }
+ \frac{1}{4} \sum_{ijkl} \bar{v}_{ijkl}  
\frac{\langle  \Psi_N | a^\dagger_i \, a^\dagger_j \, a_l \, a_k | \Psi _{N} \rangle }{\langle  \Psi_N | \Psi _{N} \rangle }. \nonumber \\
&=& {E}_{N} \left[ \rho^N , R^N \right] =   
 \sum_{ij} t_{ij} \,  \rho^N_{ij}  + \frac{1}{4} \sum_{ijkl} \bar{v}_{ijkl}   R^N_{kl, ij}. \label{eq:onetwo}
\end{eqnarray} 
Therefore, we see that the projected energy can be considered either as a functional of the transition densities (Eq. 
(\ref{eq:ekernel})) or as a functional of the one- and two-body density matrices (called hereafter OBDM and TBDM) of the projected state (Eq. (\ref{eq:onetwo})). The two formulations are strictly 
equivalent and can be connected by expressing the density matrices in terms of the gauge angles as:
 \begin{eqnarray}
\rho^{N}_{ij} =   \int_{0}^{2\pi}  d\varphi \rho^{0\varphi}_{ij}    {\cal N}_N ({0, \varphi}) , \hspace*{1.cm} 
 R^N_{kl, ij} =   \int_{0}^{2\pi}  d\varphi (\rho^{0\varphi}_{ki} \rho^{0\varphi}_{lj} -\rho^{0\varphi}_{kj} \rho^{0\varphi}_{li}
 +    \kappa^{0\varphi}_{kl} 
{\kappa_{ij}^{\varphi 0 }}^\star  )  {\cal N}_N ({0, \varphi}) 
\end{eqnarray}
Indeed, starting from Eq. (\ref{eq:ekernel}) and combining the $\rho\rho$ and the $\kappa\kappa$ terms (Eq. \ref{eq:denssrhamil}), one 
can indeed directly recognize the components of the TBDM.

\section{Particle number restoration in the EDF case}

The formulation of broken symmetry and its restoration in the nuclear EDF
closely follows the Hamiltonian case. In a first step, called Single-Reference EDF (SR-EDF), an energy written as a functional of an auxiliary quasi-particle 
 state is given. Restricting here to bilinear functional based on effective interactions, it can be written as   
 \begin{eqnarray}
 {\cal E}_{SR}[\Phi_0] =&& \mathcal{E}_{SR} \left[ \rho , \kappa, \kappa^* \right] =  \sum_{i} t_{ii} \rho_{ii}+ \frac{1}{2} \sum_{i,j} \overline{v}_{ijij}^{\rho \rho}  \rho_{ii}\rho_{jj}  
+ \frac{1}{4} 
 \sum_{i,j} \overline{v}_{i\bar\imath j\bar\jmath }^{\kappa \kappa} \kappa_{i \bar\imath }^* \kappa_{j \bar\jmath } \, .
\label{eq:denssr}
\end{eqnarray}
The EDF theory differs from the Hamiltonian case by the appearance of effective vertices 
${\bar v}^{\rho \rho}$ and $\bar v^{\kappa \kappa}$ in the mean-field and pairing channels. In particular, these 
vertices generally (i) differ from each other, (ii) might not be fully antisymmetric, (iii) might be density dependent. 

These differences are at the heart of problems occurring when trying to extend SR-EDF theory to the Multi-Reference EDF 
framework. Standard approach to perform MR-EDF consists in using equation  (\ref{eq:ekernel}) where $E_{SR}$ is replaced 
by its functional counterpart ${\cal E}_{SR}$. However, this direct mapping has led to inconsistencies: First, MR-EDF can lead to
jumps and/or divergences in the energy \cite{Ang01,Dob07,Lac09,Ben09,Dug09}. This problem originates in the strategy used to construct MR-EDF and requires a specific regularization \cite{Lac09,Ben09,Dug09} that is possible only
for specific density dependence of the energy. Secondly, the use of density dependent term, that has been so successful 
in SR-EDF is at the origin of technical and conceptual difficulties. Which density should be used in the effective kernels still needs 
to be clarified \cite{Rob10}. Finally, the direct application of Eq. (\ref{eq:ekernel}) or any equivalent for other projection 
does not insure in the EDF case the proper behavior of the energy with respect to the symmetry transformation \cite{Dug10}. 
None of these problems occur in the Hamiltonian case. This is clearly shown by expression  (\ref{eq:onetwo}), noting that projected densities have all the required properties (symmetry transformation, anti-symmetrization, non-divergence and regularity...). 

\begin{figure}[h]
\includegraphics[width=20pc]{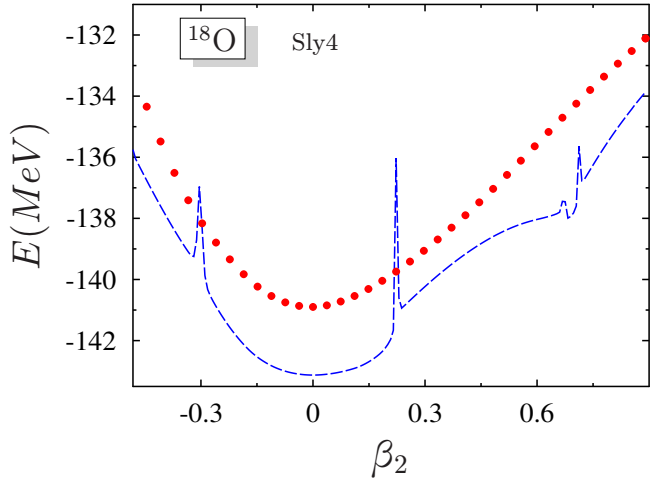}\hspace{2pc}%
\begin{minipage}[b]{14pc}\caption{\label{fig:1mbdc} 
(Color online)
Particle-number restored deformation energy surface of $^{18}$O calculated using Eq. (\ref{eq:edftotprojcor})
when the SLy4 effective interaction is used in the 
particle-hole channel. The dashed line corresponds to the non-regularized MR-EDF result 
directly obtained by gauge angle integration using 199 points in the discretization (Eq. (\ref{eq:ekernel})).
The filled circles correspond to the result obtained using the Symmetry-Conserved EDF based on
 equation (\ref{eq:edftotprojcor}).}
\end{minipage}
\end{figure}

It is therefore quite natural to see if and under which conditions, an equation similar to (\ref{eq:onetwo}) can be obtained starting from the MR-EDF approach. We have recently shown \cite{Hup11}, that starting with [a] a bilinear functional, [b] with density independent 
interaction and [c] a slight modification of the renormalization proposed in ref. \cite{Lac09}, the MR-EDF obtained by gauge angle integration is equivalent to a functional of the projected state densities written as:
\begin{eqnarray}
 {\cal E}_{N} [\Psi_{N}] &=& \sum_{i} t_{ii} n^N_i 
+ \frac{1}{2} \sum_{i,j, j\neq \bar\imath } \overline{v}_{ijij}^{\rho \rho}   R^N_{ijij} 
+ \frac{1}{4} \sum_{i \neq j, j\neq \bar\imath } \overline{v}_{i\bar\imath j\bar\jmath }^{\kappa \kappa}  R^N_{j \bar\jmath  i\bar\imath  }  
\nonumber \\
&+& \frac{1}{2} \sum_{i} \overline{v}_{i\bar\imath  i \bar\imath }^{\rho \rho}  n^N_i n^N_i 
+\frac{1}{2} \sum_{i}\overline{v}_{i\bar\imath i\bar\imath }^{\kappa \kappa}  n^N_i (1 - n^N_i)  \,  , 
 \label{eq:edftotprojcor}
\end{eqnarray}  
where $n^N_i$ denotes the occupation numbers of $| \Psi_N \rangle$. This expression clearly shows that conditions [a-c] insure
a well behaved functional as far as the particle number projection is concerned. It also shows that the MR-EDF can eventually be interpreted as a functional of densities constructed from an auxiliary state $| \Psi_N \rangle$ which leads to the functional 
sequence:
\begin{eqnarray}
\Psi_N~\Longrightarrow~ ( \rho^N , R^N) ~\Longrightarrow ~ {\cal E}_{N}.
\end{eqnarray}
The direct use of 
Eq. (\ref{eq:edftotprojcor}) instead of Eq. (\ref{eq:ekernel}) is possible and will lead to result identical to MR-EDF when the 
regularization is applicable. 
This formulation where breaking and restoration are accounted for simultaneously will be called hereafter 
Symmetry-Conserving EDF  (SC-EDF). SC-EDF can be used with density dependent interaction, including dependence in 
non-integer powers of the density, and  can be formulated as a consistent extension of the SR-EDF approach. 
Then, the density entering in the 
effective vertices becomes the density of the projected state. An illustration of the SC-EDF is shown in figure \ref{fig:1mbdc}
for a functional containing non-integer powers of the density. The MR-EDF results where no regularization is possible 
\cite{Dug10} is also shown as a reference.
\section{Summary}
The MR-EDF approach applied to particle number projection has been further analyzed here. Guided by the Hamiltonian case, we show that, under certain conditions, this approach can be formulated as a functional theory of the projected one- and 
two-body density matrices. This alternative formulation called Symmetry Conserving-EDF can eventually be applied to
functional where the MR-EDF is ill defined. Note finally that, it has also been shown recently that the two-body density matrix 
of the projected is a functional of the one-body density matrix \cite{Lac10b, Hup11b}.

\end{document}